\documentclass[english,two column, single space]{revtex4-1}
\usepackage[T1]{fontenc}
\usepackage[latin9]{inputenc}
\usepackage{color}
\usepackage{amsmath}
\usepackage{amssymb}
\usepackage{graphicx}

\makeatletter

\@ifundefined{textcolor}{}
{%
 \definecolor{BLACK}{gray}{0}
 \definecolor{WHITE}{gray}{1}
 \definecolor{RED}{rgb}{1,0,0}
 \definecolor{GREEN}{rgb}{0,1,0}
 \definecolor{BLUE}{rgb}{0,0,1}
 \definecolor{CYAN}{cmyk}{1,0,0,0}
 \definecolor{MAGENTA}{cmyk}{0,1,0,0}
 \definecolor{YELLOW}{cmyk}{0,0,1,0}
 }

\@ifundefined{definecolor}
 {\usepackage{color}}{}
\makeatother

\makeatother

\usepackage{babel}
\begin{document}

\title{Einstein-Podolsky-Rosen paradox and quantum steering in pulsed optomechanics }

\author{Q. Y. He$^{1,2}$ and M. D. Reid$^{1}$}

\affiliation{$^{1}$Centre for Quantum Atom Optics, Swinburne University of Technology,
Melbourne, Australia\\
$^{2}$State Key Laboratory of Mesoscopic Physics, School of Physics,
Peking University, Beijing 100871, China}

\begin{abstract}
We describe how to generate an Einstein-Podolsky-Rosen (EPR) paradox
between a mesoscopic mechanical oscillator and an optical pulse. We
find two types of paradox, defined by whether it is the oscillator
or the pulse that shows the effect Schrodinger called {}``steering''.
Only the oscillator paradox addresses the question of mesoscopic local
reality for a massive system. In that case, EPR's {}``elements of
reality'' are defined for the oscillator, and it is these elements
of reality that are falsified (if quantum mechanics is complete).
For this sort of paradox, we show that a thermal barrier exists, meaning
that a threshold level of pulse-oscillator interaction is required
for a given thermal occupation $n_{0}$ of the oscillator. We find
there is no equivalent thermal barrier for the entanglement of the
pulse with the oscillator, nor for the EPR paradox that addresses
the local reality of the optical system. Finally, we examine the possibility
of an EPR paradox between two entangled oscillators. Our work highlights
the asymmetrical effect of thermal noise on quantum nonlocality.
\end{abstract}
\maketitle

\section{Introduction}

It is an outstanding challenge in fundamental physics to test quantum
nonlocality for mesoscopic, massive systems. The Einstein-Podolsky-Rosen
(EPR) paradox \cite{epr} is one of the most powerful tests of quantum
nonlocality. Presented originally as an argument for the completion
of quantum mechanics (QM),  the EPR paradox has been experimentally
realised so far only in optics \cite{wu,grangierepraspect ,kim,rrmp,boyd,recenteprexp}. 

The observation of an EPR paradox for the position and momentum of
mesoscopic mechanical oscillators would represent an important advance,
since this would demonstrate the inconsistency of QM with the local
reality (LR) of a massive object. First proposed by Giovannetti et
al \cite{paulo}, such a realisation would also give an experimental
platform to probe the macroscopic reality of an object, along the
lines suggested by Schrodinger \cite{Schr=0000F6dinger-cat-1,oscil-1,Schr=0000F6dinger}.
While mesoscopic superpositions were achieved with ion-traps and microwave
oscillators, the use of nanomechanical oscillators creates mass distribution
superpositions, which tests the effects of gravity. 

In light of the exceptional importance of these developments, we examine
in this paper the limitations imposed by thermal noise for achieving
an EPR paradox in an optomechanical system. There have been numerous
proposals and studies, but mainly for the entanglement of optical
and/ or mechanical modes \textcolor{black}{\cite{genes quantumeffects,entpaulo,optentrobustent,entpaulo-1,enttwomech,optent1,optent2,optent3,optent5,optent2b,optent2c}}.
\textcolor{blue}{ }Relatively little is known about the paradox itself.
We expect that an EPR paradox for the positions and momentum of the
mechanical oscillator will be strongly masked by thermal motion. Advances
in cooling to the ground state of mesoscopic oscillators improve the
likelihood of the realisation of a massive particle EPR paradox \cite{ground-1}. 

First, let us recall the important features of the EPR paradox. The
original EPR state was an entangled state of two particles (which
we label $A$ and $B$) that have perfectly correlated positions ($X_{A}$,
$X_{B}$) and momenta ($P_{A}$, $P_{B}$). Measurements on particle
A give immediate information about either the position or momentum
of particle B. The EPR paradox arises because the assumption that
the measurements do not disturb particle B (Local Realism) would imply
a simultaneous and very precise predetermination for \emph{both} of
$X_{B}$ and $P_{B}$. No \emph{local} \emph{quantum} state of the
particle $B$ however can be consistent with such precise predetermination,
for both momentum and position. In this way, an inconsistency between
Local Realism (LR) and the completeness of quantum mechanics is established. 

The distinctive feature of the EPR paradox is that, unlike entanglement,
it is a form of quantum nonlocality in which the roles of the two
systems are \emph{asymmetrical}. In the above example, an inconsistency
of QM with LR is established for the local system B. The details about
the system A $-$ which acts only to give information about B$-$
are not so important. 

The main point of this paper is to understand how to obtain an irrefutable
discrepancy between quantum mechanics and the local reality of the
\emph{mechanical oscillator} system. We propose to do this by entangling
it with an optical pulse. We consider an idealised model, developed
by Hofer et al \cite{eprentosc}, for pulsed optomechanics on fast
time scales \cite{vann-1,quantum comm}. The model introduces only
two parameters: the squeeze parameter $r$ which is a measure of the
pulse-oscillator interaction, and $n_{0}$, the thermal occupation
number of the mechanical oscillator. Our main conclusion is that thermal
noise provides a strong, but not insuperable, barrier to the oscillator
EPR paradox. The barrier however is \emph{directional}, to prevent
{}``steering'' of the thermally-excited mechanical system, in a
sense we will explain below.

To detect the EPR paradox one must consider non-ideal states, and
it is not enough to simply prove entanglement. Suppose we use scaled
quadratures, so that we can write the Heisenberg uncertainty relation
for particle $B$ as $\Delta X_{B}\Delta P_{B}\geq1$. Then the simplest
form of an EPR paradox is realised when an inferred uncertainty relation
is {}``violated'' under the assumptions of LR, so that \cite{eprr}
\begin{equation}
E_{B|A}\equiv\Delta_{inf}X_{B}\Delta_{inf}P_{B}<1.\label{eq:eprcond}
\end{equation}
Here $\Delta_{inf}X_{B}$ and $\Delta_{inf}P_{B}$ are the uncertainties
associated with the collapsed wave functions created by local measurements
(made by {}``Alice'') on particle $A$. These uncertainties allow
her to infer either the position or momentum of particle $B$ to a
given accuracy, depending on the choice of her measurements. The realisation
of this EPR criterion (\ref{eq:eprcond}) poses a more difficult challenge
than the realisation of entanglement.

There has been a resurgence of interest in this area with new experiments
\cite{eprsteerphoton,smtuheprsteer,eprsteerloss,steer z,onewaysteer-1,parampsteer,boyd-1}
motivated by a realisation \cite{Wiseman,steve} that the paradox
is also an example of the nonlocality referred to as {}``steering''
\cite{Schr=0000F6dinger}. Steering gives a way to quantify how measurements
by Alice can collapse the wavepacket of $B$. For a paradox achieved
by condition (\ref{eq:eprcond}), we can conclude that Alice can steer
system $B$ \cite{Wiseman,cavaleprsteerineq}.   The EPR paradox
therefore is a stronger test of QM than entanglement. 

Our conclusion is that thermal noise $n_{0}$ provides a stronger
barrier to the EPR paradox than to entanglement. We identify two sorts
of EPR paradox: $E_{m|c}<1$ and $E_{c|m}<1$, where $m$ and $c$
are the oscillator and cavity field respectively. The most important
is $E_{m|c}<1$. By analysing the {}``elements of reality'' associated
with EPR's argument, we see it is this paradox which enables a test
of the mesoscopic nonlocality for the massive system. 

Specifically, we find that the thermal noise $n_{0}$ of the oscillator
induces a threshold for the pulse-oscillator interaction (as measured
by $r$) if one is to observe the paradox $E_{m|c}<1$. In the limit
of large $n_{0}$, we require $r>\frac{1}{2}\ln2$. Consistent with
the fact that the field is not thermally excited, we find there is
no similar thermal barrier for an EPR paradox $E_{c|m}<1$, which
demonstrates a {}``steering'' of the optical system. The oscillator-pulse
system therefore exhibits regimes of {}``one-way'' steering \cite{onewaysteer-1,murrayoneway},
where only the steering of the pulse is detectable. 

We also find there is no (similar) thermal barrier for the entanglement
between the optical pulse and oscillator. In this dissipation-free
model, entanglement can exist for any $n_{0}$ and $r>0$. We will
see however that the thermally insensitive entanglement must manifest
in an \emph{asymmetric} way, by measurement of the variances of quantities,
$X_{A}-g_{x}X_{B}$, $P_{A}+g_{p}P_{B}$, where $g_{x}$,$g_{p}$
are selected real numbers, not equal to $1$. 

As a final result of the paper, we examine the possibility of an EPR
paradox between two mechanical oscillators that are thermally excited.
This leads us to distinguish a subclass of {}``symmetric'' entanglement,
which can be detected with $g_{x}=g_{p}=1$, and for which a thermal
barrier $r>\frac{1}{2}\ln n_{0}$ \emph{does} exist. We are able to
show that this symmetric form of entanglement is relevant to the creation
of entanglement between two symmetric thermal oscillators, and therefore
has its own fundamental significance. The symmetric entanglement is
detected by the criterion of Duan et al \cite{duan-1}. By examining
two asymmetrically excited oscillators, we conclude that the {}``EPR
steering'' of one by the other can be made largely insensitive to
the level of thermal excitation of one of the oscillators.

\begin{figure}[b]
\begin{raggedright}
\includegraphics[width=0.8\columnwidth]{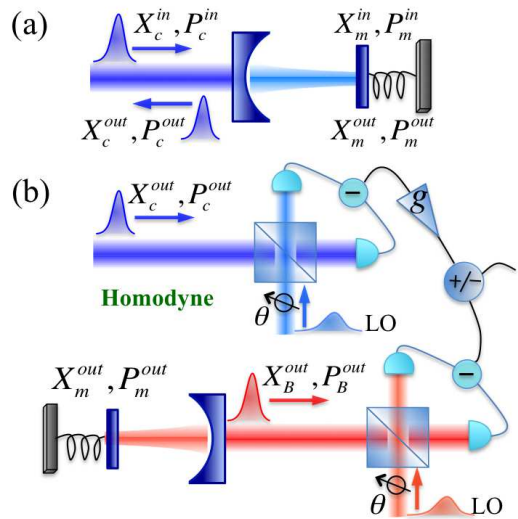}
\par\end{raggedright}

\caption{\textcolor{black}{(Color online) }Measurement of the EPR paradox between
an oscillator and a pulse\textcolor{black}{\emph{. (a) Entangling
a pulse with a mechanical oscillator. }}\textcolor{black}{Following
HWAH, an {}``entangling'' blue-detuned pulse interacts with an optomechanical
system. The output pulse amplitudes $X_{c}^{out}$, $P_{c}^{out}$
are EPR correlated with the final quadratures $X_{m}^{out}$, $P_{m}^{out}$
of the mechanical oscillator, according to $X_{c}^{out}\sim-P_{m}^{out}$
and $P_{c}^{out}\sim-X_{m}^{out}$, in the limit of a large squeezing
parameter $r$. }\textit{\textcolor{black}{\emph{(b)}}}\textit{ To
verify the EPR paradox.}\textcolor{black}{{} The output pulse amplitudes
$X_{c}^{out}$, $P_{c}^{out}$ are measured by homodyne detection.
The quadratures of the oscillator can be measured by interacting the
cavity with a second {}``verifying'' red-detuned pulse.}\label{fig:oscillator and pulse} }
\end{figure}

\section{The HAWH Model}

We consider a mechanical oscillator coupled to an optical mode of
a cavity \cite{paulo}. Hofer, Wieczorek, Aspelmeyer and Hammerer
(HWAH) \cite{eprentosc} proposed a scheme (Figure \ref{fig:oscillator and pulse})
in which a light pulse is input to an optomechanical cavity mode and
interacts with the oscillator mirror mode via radiation pressure \cite{mantomb}.
The pulse emerges from the cavity with quadratures that are EPR correlated
with those of the oscillator. The effective interaction Hamiltonian
\cite{mantomb,prod osc,cklaw} for the cavity-oscillator system in
a frame rotating at the laser frequency is 
\begin{equation}
H=\omega_{m}a_{m}^{\dagger}a_{m}+\Delta_{c}a_{c}^{\dagger}a_{c}+g_{R}(a_{m}+a_{m}^{\dagger})(a_{c}+a_{c}^{\dagger}),
\end{equation}
where $\Delta_{c}=\omega_{c}-\omega_{1}$ is the detuning of the cavity
with respect to the laser \cite{eprentosc}. The boson creation and
destruction operators for the optical cavity and mechanical modes
are $a_{c}$, $a_{c}^{\dagger}$ and $a_{m}$, $a_{m}^{\dagger}$
respectively. Quadrature phase amplitudes $X_{c/m}$, $P_{c/m}$ are
defined according to $a_{c}=(X_{c}+iP_{c})/2$ and $a_{m}=(X_{m}+iP_{m})/2$,
where the choice of scaling ensures the normalised EPR inequality
$\Delta_{inf}X_{B}\Delta_{inf}P_{B}<1$. The term in $g_{R}$ describes
the linearised optomechanical coupling due to the radiation pressure,
and comprises both a beam splitter-type coupling (involving $a_{m}^{\dagger}a_{c}+a_{c}^{\dagger}a_{m}$)
and a two-mode squeezing interaction term (involving $a_{m}a_{c}+a_{m}^{\dagger}a_{c}^{\dagger}$)
of the type known to generate EPR entanglement \cite{eprr,eprtwomode}. 

The physical parameters of the HAWH pulse-oscillator model are the
interaction strength $g_{R}$, the oscillation frequency $\omega_{m}$
, with dissipation rate $\gamma$, the optical cavity resonance frequency
$\omega_{c}$ with decay rate $\kappa$, the pulse carrier frequency
$\omega_{1}$ and duration time $\tau$. The initial occupation number
$n_{0}$ of the thermal state of the mirror is a vital number which
determines the `quantumness' of the system. HWAH propose the pulse
to be either blue-detuned or red-detuned to the cavity resonance \cite{eprentosc},
to enhance either the two-mode squeezing term (for the purpose of
generating entanglement) or the beam splitter-type term (for the purpose
of measurement).

To generate the correlations of the EPR paradox, a blue-detuned pulse
interacts with the oscillator cavity. In the case where $g_{R}\ll\kappa\ll\omega_{m}$,
HWAH derive a set of idealised Langevin equations for the mode operators.
To justify neglecting decoherence, they assume the pulse duration
and it's interaction time with the cavity are short compared to the
mechanical decoherence time. The effect of the coupling of the oscillator
to an environmental heat bath is ignored.

For the blue-detuned pulse, after making a rotating wave approximation
(RWA), with an adiabatic solution for the cavity mode, the simplified
Langevin equations lead to solutions for quadratures $X_{c}^{out}$,
$P_{c}^{out}$. The solutions are \cite{eprentosc,inputoutput}
\begin{eqnarray}
X_{c}^{out} & = & -e^{r}X_{c}^{in}-\sqrt{e^{2r}-1}P_{m}^{in},\nonumber \\
P_{c}^{out} & = & -e^{r}P_{c}^{in}-\sqrt{e^{2r}-1}X_{m}^{in},\nonumber \\
X_{m}^{out} & = & e^{r}X_{m}^{in}+\sqrt{e^{2r}-1}P_{c}^{in},\nonumber \\
P_{m}^{out} & = & e^{r}P_{m}^{in}+\sqrt{e^{2r}-1}X_{c}^{in},\label{eq:pulseoutput}
\end{eqnarray}
where $X_{m}^{out}$ and $P_{m}^{out}$ are the final quadratures
of the mechanical oscillator, and $r=g_{R}^{2}\tau/\kappa$ is the
{}``squeezing parameter''. The initial quadratures of the oscillator
incorporate the effect of the thermal excitation parameter $n_{0}$. 

The input-output solutions (\ref{eq:pulseoutput}) are similar to
those of a two-mode squeezed state \cite{caves sch-1} and will form
the basis for modeling the fundamental constraints provided by the
thermal noise for an EPR paradox. The solutions (\ref{eq:pulseoutput})
in the limit of large $r$ become $X_{c}^{out}=-e^{r}(X_{c}^{in}+P_{m}^{in})$
and $P_{c}^{out}=-e^{r}(P_{c}^{in}+X_{m}^{in})$. The EPR nature of
the correlations is evident, since 
\begin{eqnarray}
X_{m}^{out} & = & -P_{c}^{out},\ P_{m}^{out}=-X_{c}^{out}
\end{eqnarray}
so that a measurement of the quadrature $X_{c}^{out}$ (or $P_{c}^{out}$)
of the pulse will immediately give the prediction for the quadrature
$-P_{m}^{out}$ (or $-X_{m}^{out}$) of the oscillator. 

The HAWH model is very idealised, and further work is needed to test
the validity of the approximations for the pulse and to model the
significant decoherence expected for an oscillator interacting with
its environment. The model does however capture the main physical
effects that generate an EPR correlation, and gives a treatment of
the thermal noise of the initial state of the oscillator. It can be
therefore be used to give a first order understanding of the asymmetrical
effects of thermal noise on the EPR correlation, and of the level
of thermal cooling that may be necessary, in order to observe an EPR
paradox.

\section{Detecting the Entanglement\textbf{\emph{ }}}

Often, entanglement is measured as a reduction in two variances, $\{\Delta(X_{m}^{out}+P_{c}^{out})\}^{2}$
and $\{\Delta(P_{m}^{out}+X_{c}^{out})\}^{2}$ that involve symmetric
weightings of oscillator and field quadratures \cite{duan-1}. The
symmetric criterion of Duan, Giedke, Cirac and Zoller (DGCZ) \cite{duan-1}
detects entanglement when 
\begin{equation}
\{\Delta(X_{m}^{out}+P_{c}^{out})\}^{2}+\{\Delta(P_{m}^{out}+X_{c}^{out})\}^{2}<4\label{eq:duan}
\end{equation}
where we denote the variance using the notation $\{\Delta x\}^{2}\equiv\langle x^{2}\rangle-\langle x\rangle^{2}$.
This criterion, however, is far from being an optimal signature for
entanglement, owing to intrinsic asymmetries. 

Here, we examine a less restrictive condition. Entanglement between
the oscillator and pulse is proved if one can show \cite{PPT-simon,proof for product form}
\begin{equation}
\Delta_{g,ent}=\frac{\{\Delta(X_{m}^{out}+g{}_{x}P_{c}^{out})\}^{2}\{\Delta(P_{m}^{out}+g{}_{p}X_{c}^{out})\}^{2}}{\left[|g_{x}g_{p}|+1\right]^{2}}<1,\label{eq:D_product}
\end{equation}
where $g_{x}$ and $g_{p}$ are arbitrary real numbers. The variances
in the numerator are directly measurable by the scheme depicted in
Fig. \ref{fig:oscillator and pulse}, where the $g_{x}$ and $g_{p}$
are classical gain factors. Here $\Delta_{g,ent}$ can be minimized
to a value $\Delta_{ent}$ by choosing the optimal factor $g_{x}=g_{p}=g$
\begin{align}
g & =\frac{\delta+\sqrt{\delta^{2}+4e^{2r}(e^{2r}-1)}}{2e^{r}\sqrt{e^{2r}-1}},\label{eq:gopt}
\end{align}
where $\delta=\frac{\Delta^{2}X_{m}^{in}-\Delta^{2}P_{c}^{in}}{\Delta^{2}X_{m}^{in}+\Delta^{2}P_{c}^{in}}$.
We assume the initial state of the light field to be the vacuum state
and that of the mirror to be a thermal state with mean excitation
number $n_{0}$. In this case, $\delta=\frac{n_{0}}{n_{0}+1}$.  

It is shown elsewhere \cite{discord,duan-1} that the condition given
by (\ref{eq:D_product}) with optimal $g$ is equivalent to the positive
partial transpose (PPT) condition developed by Simon \cite{PPT-simon}
for Gaussian states, and is therefore necessary and sufficient for
Gaussian two-mode entanglement. The entanglement $\Delta_{ent}$
can be measured by the arrangement of Figure \ref{fig:oscillator and pulse}. 

\begin{figure}[t]

\begin{raggedright}
\includegraphics[width=0.9\columnwidth]{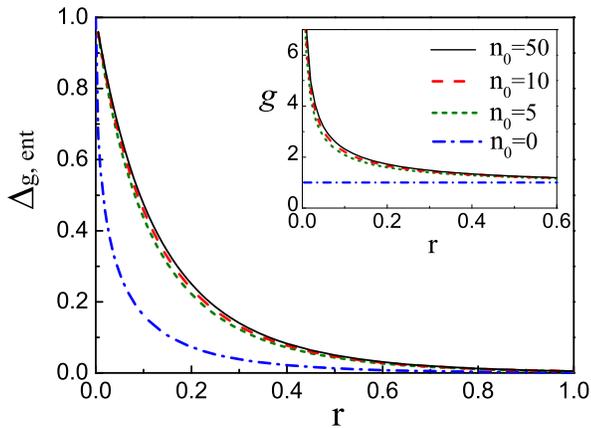}
\par\end{raggedright}

\centering{}\caption{\textcolor{black}{(Color online) Entanglement} $\Delta_{g,ent}$ plotted
versus the squeezing parameter $r$ for $n_{0}=0,\ 5,\ 10,\ 50$,
where $n_{0}$ is initial occupation number of the oscillator. Entanglement
between the oscillator and pulse is observed when $\Delta_{ent}<1$.
Strong entanglement occurs when $\Delta_{ent}\rightarrow0$. The optimal
$g$ to minimize $\Delta_{g,ent}$ is shown in the inset. The results
indicate presence of entanglement, even for large $n_{0}$. \label{fig:asymmetric_ent_sum_product}
\textcolor{red}{ }}
\end{figure}

We note that, unlike {}``steering'', entanglement is defined as
a property that the two systems share without specification of direction
$-$ that is, if $A$ is entangled with $B$, we know that $B$ is
entangled with $A$. Consistent with this, we see that the criterion
(\ref{eq:D_product}) is symmetric with respect to interchange of
$m$ and $c$, provided $g_{x}$ and $g_{p}$ are interchanged with
their reciprocals. Thus, entanglement can be detected using either
criterion (\ref{eq:D_product}), or the criterion obtained in interchanging
$m$ with $c$, provided the choice of the $g_{x}$'s and $g_{p}$'s
is kept fully flexible. The prediction of the model (\ref{eq:pulseoutput})
for the entanglement measured by $\Delta_{g,ent}$ is plotted in Fig.
\ref{fig:asymmetric_ent_sum_product}. 

Now we come to the first important result of this paper. Surprisingly,
we see from the Fig. \ref{fig:asymmetric_ent_sum_product} that no
thermal barrier exists for the presence of entanglement. For any given
thermal occupation number $n_{0}$, we can always show entanglement
for $r>0$. \textcolor{black}{}In other words, once the system is
entangled, no amount of thermal noise can completely destroy it. A
similar type of robustness of entanglement has been predicted for
the steady state opto-mechanical entanglement that is generated using
continuous wave light fields \cite{entpaulo,entpaulo-1,genes quantumeffects,optentrobustent}.
We note from the Figure that the detection of this \emph{thermally
insensitive} entanglement is linked to values of the parameter $g_{opt}$
that are very different to $1$. As we will see below, this result
can be understood in terms of the concept of quantum steering. 

This result contrasts with that obtained for entanglement detected
using the \emph{symmetric} DGCZ condition (\ref{eq:duan}) \cite{duan-1},
given by $\Delta_{g,ent}<1$ where $g_{x}=g_{p}=1$. We call this
sort of entanglement {}``symmetric entanglement''. In that case,
a thermal barrier does exist, and entanglement can be detected only
when the squeezing parameter is sufficiently large \cite{eprentosc},
\begin{equation}
r>\ln\frac{n_{0}+2}{2\sqrt{n_{0}+1}}\overset{n_{0}\rightarrow\infty}{\sim}\frac{1}{2}\ln n_{0}.\label{eq:r_dgcz}
\end{equation}
While this thermal barrier becomes relevant to detecting entanglement
between two symmetric, thermally excited oscillators, it does not
place a limit on the detection of entanglement between the oscillator
and a pulse.

\section{Detecting an EPR paradox and quantum steering}

\subsection{Quantum steering of the mechanical oscillator by the pulse}

Now we examine how to detect an EPR paradox. An EPR paradox is confirmed
if \cite{eprr} 
\begin{equation}
E_{m|c}=\Delta_{inf}X_{m}^{out}\Delta_{inf}P_{m}^{out}<1,\label{eq:steering-1}
\end{equation}
where $\Delta_{inf}X_{m}^{out}$ is the error in the prediction for
the value of the oscillator's position, made by a measurement on the
pulse. The $\Delta_{inf}P_{m}^{out}$ is defined similarly. The realisation
of $E_{m|c}<1$ is verification of a quantum steering of the mechanical
system by measurements made on the pulse \cite{Wiseman,cavaleprsteerineq}.
For the subclass of quantum systems given by Gaussian states and measurements,
as is the case here, this criterion becomes necessary and sufficient
to detect steering of the system $m$ by the second system \cite{Wiseman}.
Walborn et al have derived a more sensitive entropic criterion for
{}``EPR steering'' that is useful in other cases \cite{entropic epr criterion}.

A simple way to determine the conditional uncertainties for Gaussian
distributions is to use a linear estimate $g{}_{x}P_{c}^{out}$, based
on the result $P_{c}^{out}$ for measurement at $A$ \cite{eprr,rrmp}.
We find
\begin{eqnarray}
\Delta_{inf}X_{m}^{out} & = & \Delta(X_{m}^{out}-g_{x}P_{c}^{out}),
\end{eqnarray}
where $g{}_{x}=\langle X_{m}^{out},P_{c}^{out}\rangle/\{\Delta P_{c}^{out}\}^{2}$
is optimised to minimise $\{\Delta_{inf}X_{m}^{out}\}^{2}$. Here
we use the notation $\langle x,y\rangle\equiv\langle xy\rangle-\langle x\rangle\langle y\rangle$.
Similarly, the conditional variance $\{\Delta_{inf}P_{m}^{out}\}^{2}$
is evaluated by
\begin{eqnarray}
\{\Delta_{inf}P_{m}^{out}\}^{2} & = & \{\Delta(P_{m}^{out}+g_{p}X_{c}^{out})\}^{2},
\end{eqnarray}
where $g{}_{p}=\langle P_{m}^{out},X_{c}^{out}\rangle/\{\Delta X_{c}^{out}\}^{2}$.
We note that the values of $g_{x}$, $g_{p}$ that optimise for the
EPR paradox are generally different to those that optimise the entanglement
given by (\ref{eq:D_product}). 

We assume the light to be initially in a vacuum state and the mechanical
oscillator to be initially in a thermal state, with mean occupation
number $n_{0}$. We can then calculate the prediction of the model
(\ref{eq:pulseoutput}) for the EPR paradox. We find 
\begin{eqnarray}
\Delta_{inf}X_{m}^{out} & = & \Delta_{inf}P_{m}^{out},
\end{eqnarray}
where we take $g_{x}=g_{p}=g$ and 
\begin{equation}
g=\frac{2e^{r}\sqrt{e^{2r}-1}(n_{0}+1)}{2e^{2r}(n_{0}+1)-(2n_{0}+1)}.\label{eq:epr(m|c)_g}
\end{equation}

The EPR paradox parameter $E_{m|c}$ is given in the Fig. \ref{fig:steering_one_way_two_way}
versus $r$, for various values of initial thermal occupation $n_{0}$.
The EPR paradox realised when $E_{m|c}<1$ is predicted for values
of squeezing parameter given by 
\begin{equation}
r>r_{epr}=\frac{1}{2}\ln\frac{2n_{0}+1}{n_{0}+1}\overset{n_{0}\rightarrow\infty}{\sim}\frac{1}{2}\ln2.\label{eq:r_epr}
\end{equation}
There is a temperature-dependent minimal squeezing parameter $r_{epr}$
required to observe the paradox. We note however that for large thermal
excitation $n_{0}$, the barrier becomes fixed, at $r_{epr}=\frac{1}{2}\ln2$
as $n_{0}\rightarrow\infty$. 

This is the second noteworthy result. If the squeeze parameter $r$
is large enough (that is, if there is enough entanglement between
the oscillator and the pulse), the quantum steering of the oscillator
by the pulse cannot be destroyed by further increasing the thermal
noise value $n_{0}$. \textcolor{red}{}The quantum steering of the
oscillator by the pulse can be achieved when $r>\frac{1}{2}\ln2$.\textcolor{black}{{}
We expect that this effect arises because the second EPR system, the
pulse, is }\textcolor{black}{\emph{not }}\textcolor{black}{thermally
excited. A different effect is expected for the EPR paradox between
two oscillators.}

\subsection{Quantum steering of the pulse by the oscillator}

\begin{figure}[t]
\begin{centering}

\par\end{centering}

\begin{raggedright}
\includegraphics[width=0.9\columnwidth]{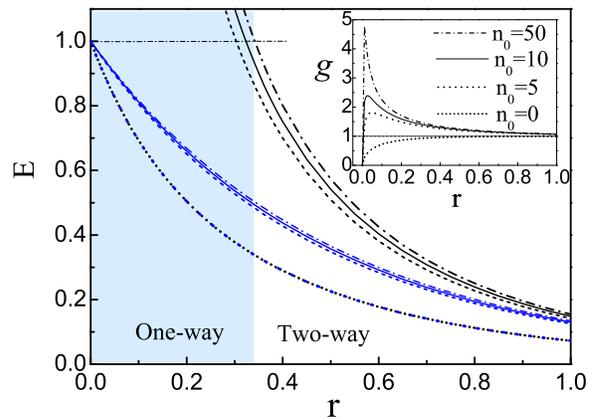}
\par\end{raggedright}

\caption{(Color online) EPR paradox and quantum steering between the oscillator
and the pulse. An EPR paradox and quantum steering of the oscillator
by the pulse is detected when $E=E_{m|c}<1$ (black upper set of lines,
that intercept with the line $E=1$ at the value for $r$ given by
$r_{epr}$ of Eq. (\ref{eq:r_epr})). An EPR paradox and the steering
of the pulse by the oscillator is detected when $E=E_{c|m}<1$ (blue
lower set of lines). \textcolor{black}{The lowest two superposed curves
show $E_{m|c}$ and $E_{c|m}$ with $n_{0}=0$ (dotted). The inset
shows the optimal $g$ to minimize }$E_{m|c}$\textcolor{black}{.\label{fig:steering_one_way_two_way} }}
\end{figure}

An EPR paradox can be shown \emph{the other way}, by the criterion
\begin{equation}
E_{c|m}=\Delta_{inf}X_{c}^{out}\Delta_{inf}P_{c}^{out}<1.\label{eq:steering from mirror}
\end{equation}
 In this case, the optical pulse is {}``steered'' by the measurements
made on the mechanical oscillator. Results for the prediction of $E_{c|m}$
based on the model (\ref{eq:pulseoutput}) are shown in Fig. \ref{fig:steering_one_way_two_way}.
Such an EPR paradox is thermally insensitive, being possible for any
value of initial thermal noise $n_{0}$, and for any squeezing parameter
$r>0$. Since EPR steering requires entanglement \cite{Wiseman},
this property underpins the thermal insensitivity noted above for
entanglement. However, because the thermally insensitive steering
is {}``one-way'' only, it does not correspond to an entanglement
that can be detected symmetrically with respect to the oscillators
$-$that is, with $g=1$. 

We note that there are two regimes for the observation of quantum
steering and the EPR paradox. For $r\leq\frac{1}{2}\ln2$, the only
EPR paradox possible is $E_{c|m}<1$ ({}``one-way steering'' \cite{steer thmurray,onewaysteer-1,steer z,eprsteerloss}).
For $r>\frac{1}{2}\ln2$, {}``two-way steering'' becomes possible,
and both paradoxes $E_{m|c}<1$ or $E_{c|m}<1$ can be confirmed.

Now we can understand the reason for the reduced sensitivity of the
entanglement to the thermal noise of the oscillator. We have shown
that a threshold squeezing parameter $r$ is necessary to enable a
steering of the thermal oscillator by the measurements made on the
pulse, but there is no threshold for the steering of the pulse by
the measurements on the oscillator. Entanglement is a defined as
a shared quantity, and must exist between the two systems if either
form of steering is achieved \cite{Wiseman}. Hence, entanglement
is detected without the threshold, because this is possible for the
quantum steering ($E_{c|m}<1$) in one direction. \textcolor{red}{}

\textcolor{black}{We argue however that the symmetric form of entanglement
has its own special significance. The DGCZ criterion (\ref{eq:duan})
is defined as that entanglement detected in a symmetric way, with
$g_{x}=g_{p}=1$. This distinguishes it from the thermally insensitive
entanglement that is detected with $g_{x}$, $g_{p}$ values very
different to $1$ (Fig. \ref{fig:steering_one_way_two_way}). For
symmetric systems, like two equivalently thermally excited oscillators,
we conjecture that the limitations for entanglement are determined
by the DGCZ criterion. }

\section{Fundamental significance of the steering of the oscillator}

To understand the importance of the EPR paradox that demonstrates
quantum steering of the oscillator by the pulse, we first review the
meaning of an EPR paradox \cite{epr}. An EPR paradox arises because
the assumption of local realism (LR) would imply a simultaneous, precise
predetermination for both of $X_{m}^{out}$ and $P_{m}^{out}$. No
quantum state of the oscillator however can be consistent with such
a predetermination, for both momentum and position. Thus, EPR argue
{}``elements of reality'' exist that cannot be described by quantum
mechanics, and an inconsistency between LR and the completeness of
quantum mechanics is revealed. 

We have found there is a thermal barrier for the EPR paradox that
corresponds to the {}``steering'' of the oscillator system. This
will make this sort of paradox more difficult to observe in practice.
We note however that there is a fundamental significance in observing
this sort of EPR paradox. On examining the EPR argument, we see that
the EPR paradox (in this case) is based on the premise that the action
of measuring the pulse cannot change the state of the oscillator.
Hence, {}``elements of reality'' are deduced for the oscillator
system (not the pulse). These {}``elements of reality'' become inconsistent
with quantum mechanics when $E_{m|c}=\Delta_{inf}X_{m}\Delta_{inf}P_{m}<1$.
Hence, if local realism LR is correct, the \emph{oscillator} cannot
be described quantum mechanically. Alternatively, with an assumption
that quantum mechanics is complete, it is the local reality of the
o\emph{scillator} that is disproved. For the second type of paradox,
it is the reality of the \emph{optical} state that is addressed. This
is less useful for direct insights about quantum effects with matter.

\section{Entanglement and EPR paradox between two mechanical oscillators}

\textcolor{red}{}A further challenge is to understand the thermal
limits for obtaining an EPR paradox between two thermally excited
mechanical oscillators. Bipartite entanglement between two mechanical
oscillators can be achieved, in principle, by {}``swapping'' the
entanglement between the oscillator $m1$ and the output pulse, to
an entanglement between the oscillator $m1$ at $A$ and a second
mechanical oscillator $m2$ at location $B$. Other methods are possible,
the simplest being to couple two cavities to two incoming entangled
light fields \textcolor{black}{\cite{enttwomech,optent1,optent2,optent3,optent5,optent2b,optent2c}}.
Most previous calculations have been limited to the generation of
entanglement in steady state regimes. 

\begin{figure}[t]
\raggedright{}\includegraphics[width=0.9\columnwidth]{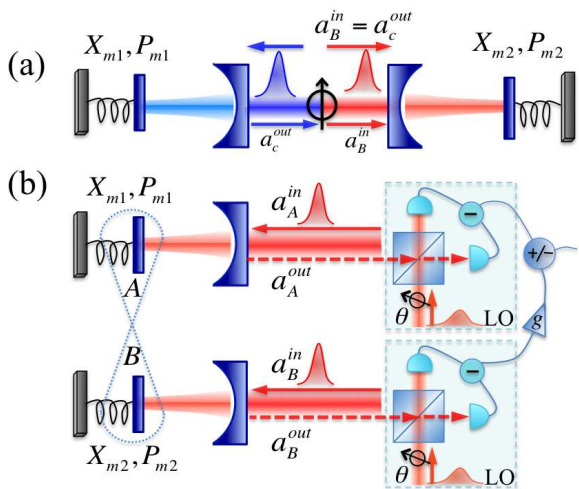}\caption{(Color online) Entangling two oscillators: (a) Generation of the entanglement
takes place when the output of the first cavity is injected into the
second cavity, as red-detuned. The final states of the two oscillators
at $A$ and $B$ will be entangled. (b) The entanglement can be verified,
at a later stage, using two red detuned pulses, and the homodyne scheme
set up as depicted, to measure the conditional inference variances
$\{\Delta_{inf}X_{mj}\}^{2}$ and $\{\Delta_{inf}P_{mj}\}^{2}$.\label{fig:two-oscillaltors-2} }
\end{figure}

In this paper, we focus for simplicity on the results of calculations
based on the first method (Fig. \ref{fig:two-oscillaltors-2}). After
interaction with the first cavity, the entangling pulse {}``carries''
the information about the quadratures of the first oscillator. As
$r\rightarrow\infty$, we see from (\ref{eq:pulseoutput}) that $X_{m1}^{out}=-P_{c}^{out}$
and $P_{m1}^{out}=-X_{c}^{out}$. Suppose then that after the coupling
to the first cavity and oscillator $m1$, the output entangling pulse
is then red-detuned relative to a second mechanical oscillator  $(m2)$
and cavity system. After an interaction with this pulse, the final
amplitudes of the second oscillator are \cite{eprentosc} 
\begin{align}
X_{m2}^{out} & =e^{-r'}X_{m2}^{in}+\sqrt{1-e^{-2r'}}P_{c}^{out},\nonumber \\
P_{m2}^{out} & =e^{-r'}P_{m2}^{in}-\sqrt{1-e^{-2r'}}X_{c}^{out},\label{eq:m2}
\end{align}
where $r'$ is the squeezing parameter of the second cavity, and $P_{c}^{out}$,
$X_{c}^{out}$ are given by (\ref{eq:pulseoutput}\textcolor{black}{).
As $r'\rightarrow\infty$, the relations between the quadratures of
the second mechanical oscillator and the pulse are $X_{m2}^{out}=P_{c}^{out}$
and $P_{m2}^{out}=-X_{c}^{out}$, which will {}``swap'' the EPR
correlation $\Delta\left(P_{c}^{out}+gX_{m1}^{out}\right)$, $\Delta\left(X_{m1}^{out}-gP_{c}^{out}\right)$
into an EPR correlation $\Delta\left(X_{m2}^{out}+gX_{m1}^{out}\right)$,
$\Delta\left(X_{m1}^{out}-gX_{m2}^{out}\right)$$ $ between the mechanical
oscillators. T}hus, an EPR parado\textcolor{black}{x }between between
the pulse and the first mechanical oscillator is directly transformed
into an EPR paradox between two mechanical oscillators in the limit
of $r'\gg r$ and $r'\rightarrow\infty$. 

\textcolor{black}{For practical reasons, since the thermal noise on
the second oscillator can be significant, it is also informative to
consider definite predictions for $r=r'$. }The final entanglement
and EPR paradox variances are readily calculated. Solving, we find
\begin{eqnarray}
\{\Delta_{inf}X_{m2}^{out}\}^{2} & = & \{\Delta\left(X_{m2}^{out}+gX_{m1}^{out}\right)\}^{2}\nonumber \\
 & = & e^{-2r}\{\Delta X_{m2}^{in}\}^{2}+(g-1)^{2}\left(e^{2r}-1\right)\{\Delta P_{c}^{in}\}^{2}\nonumber \\
 &  & \,\,\,\,+\left[\left(g-1\right)e^{r}+e^{-r}\right]^{2}\{\Delta X_{m1}^{in}\}^{2}\nonumber \\
 & = & e^{-2r}(2n_{m2}+1)+(g-1)^{2}\left(e^{2r}-1\right)\nonumber \\
 &  & \,\,\,\,+\left[\left(g-1\right)e^{r}+e^{-r}\right]^{2}(2n_{m1}+1),\label{eq:sol}
\end{eqnarray}
and $\Delta_{inf}P_{m2}^{out}=\Delta\left(P_{m2}^{out}-gP_{m1}^{out}\right)=\Delta_{inf}X_{m2}^{out}$.
Here, $n_{m1}$, $n_{m2}$ are the thermal occupation numbers for
the two oscillators.

\subsection{Entanglement }

Importantly, we note that a thermal barrier exists for the entanglement
between two oscillators with equal thermal noise $n_{m1}=n_{m2}=n_{0}$.
We examine the predictions for the entanglement criterion (\ref{eq:D_product})
but as applied to the two oscillators $m1$ and $m2$. In this case,
the optimal $g$ for the detection of entanglement (given by $\frac{\partial\Delta_{g,ent}}{\partial g}=0$)
is
\begin{eqnarray}
{\color{black}g} & {\color{black}=} & \sqrt{1+(2n_{0}+1)^{2}/4e^{4r}(n_{0}+1)^{2}}\nonumber \\
 &  & {\color{black}-(2n_{0}+1)/2e^{2r}(n_{0}+1)},\label{eq:gent}
\end{eqnarray}
which becomes $g\rightarrow1$ in the limit of large $n_{0}$ and
$r$. The threshold $ $squeezing parameter for entanglement becomes
$r_{ent}=\frac{1}{2}\ln2n_{0}$ in this limit, which has the same
large $n_{0}$ dependence as for the symmetric entanglement that is
detected by the DGCZ entanglement criterion. The similarity is expected
for two equivalent oscillators, since any entanglement that can be
created between the two oscillators must be symmetric i.e. unchanged
on interchange $A\longleftrightarrow B$, which will require $g_{opt}=g_{x}=g_{p}=1$.

\subsection{EPR paradox and steering }

\begin{figure}[t]
\begin{centering}
\includegraphics[width=0.9\columnwidth,height=4cm]{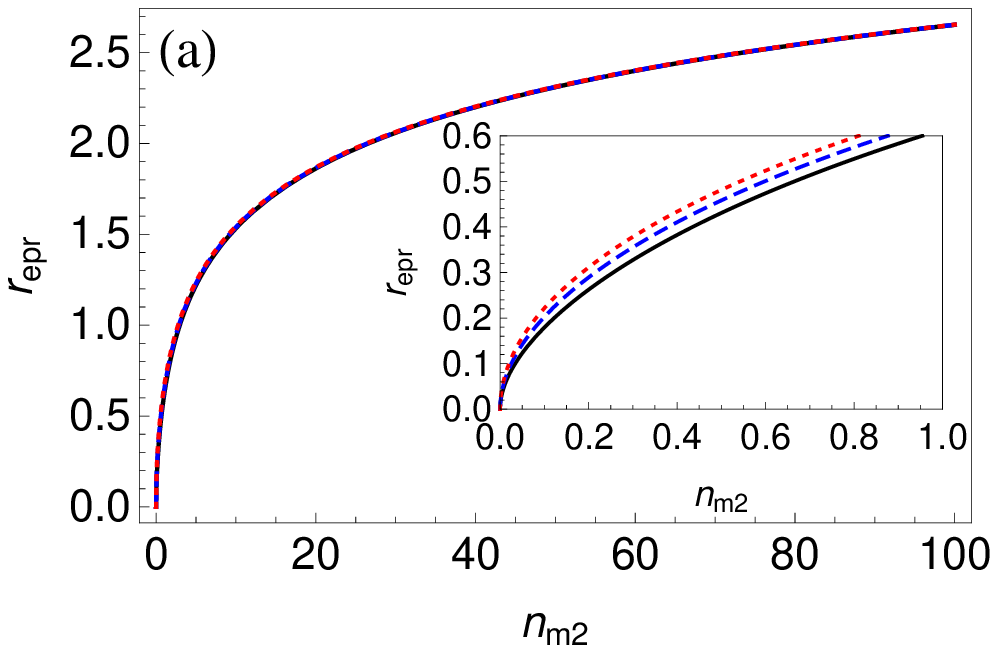}
\par\end{centering}

\begin{centering}
\includegraphics[width=0.9\columnwidth,height=4cm]{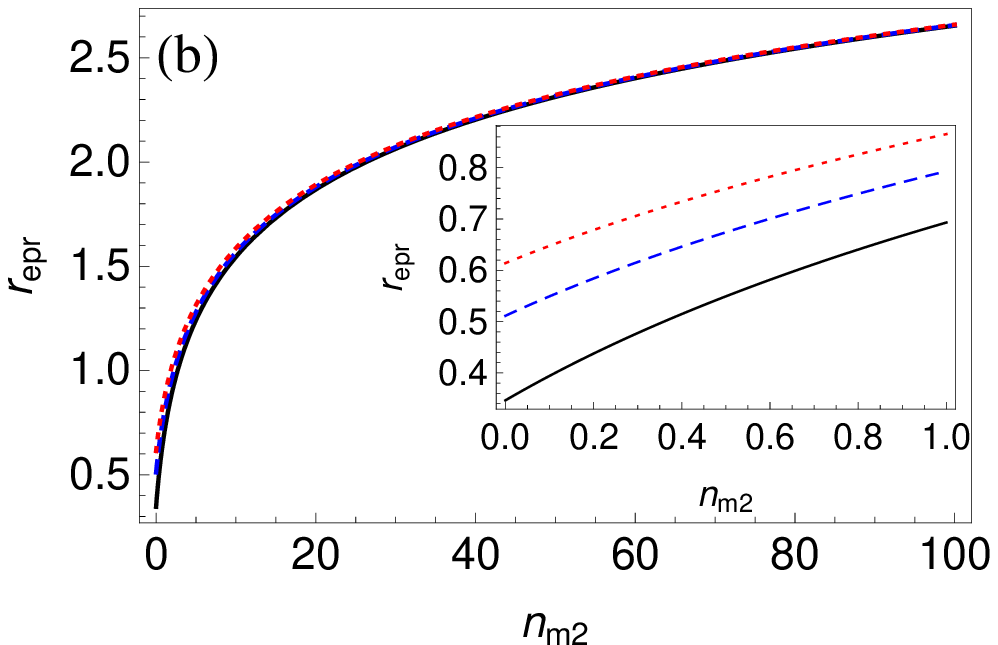}
\par\end{centering}

\begin{centering}
\includegraphics[width=0.9\columnwidth,height=4cm]{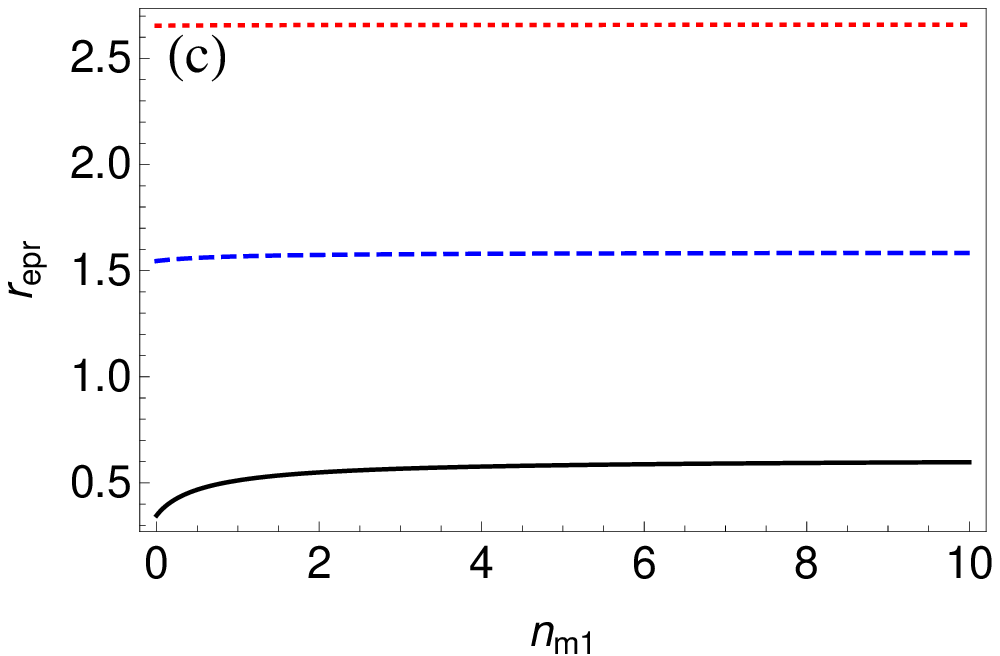}
\par\end{centering}

\centering{}\caption{(Color online) Detecting an EPR paradox between two mechanical oscillators
prepared using the scheme of Eqs. (\ref{eq:m2}) with $r=r'$: (a)
The threshold squeeze parameter $r_{epr}$ for observation of the
EPR paradox $E_{m2|m1}<1$, versus the thermal occupation number of
the oscillator $m2$. The curves are for $n_{m1}=0$ (solid), $n_{m1}=1$
(dashed), $n_{m1}=1.5\times10^{6}$ (dotted) \textcolor{red}{}(b)
The threshold squeeze parameter $r_{epr}$ for observation of the
EPR paradox $E_{m1|m2}<1$, versus the thermal occupation number of
the oscillator $m2$. The curves are for $n_{m1}=0$ (solid), $n_{m1}=1$
(dashed), $n_{m1}=1.5\times10^{6}$(dotted). \textcolor{red}{}\textcolor{black}{{}
(c)} The threshold squeeze parameter $r_{epr}$ for observation of
the EPR paradox $E_{m1|m2}<1$, versus the thermal occupation number
of the oscillator $m1$. The curves are for $n_{m2}=0$ (solid), $n_{m2}=10$
(dashed), $n_{m2}=100$ (dotted)\label{fig:rth}}
\end{figure}

We now consider the EPR paradox and quantum steering, in particular
how the oscillator $m1$ {}``steers'' the oscillator $m2$. The
optimal $g$ for the detection of the EPR paradox $E_{m2|m1}$ is
given by $\frac{\partial\Delta_{inf}X_{m2}^{out}}{\partial g}=0$
. Solving gives
\begin{eqnarray}
g & = & \frac{\left(e^{2r}-1\right)\left(n_{m1}+1\right)}{e^{2r}\left(n_{m1}+1\right)-1/2}.\label{eq:gepr1}
\end{eqnarray}
The threshold squeezing parameter for the EPR paradox $E_{m2|m1}$
is then
\begin{eqnarray}
 & r_{epr}= & \frac{1}{2}\ln\Bigl(n_{m2}+1+\sqrt{\left(n_{m2}+1\right)^{2}-\frac{n_{m1}+n_{m2}+1}{n_{m1}+1}}\Bigr).\nonumber \\
\label{eq:rth}
\end{eqnarray}
\textcolor{black}{The threshold $r_{epr}$ is plotted in Fig. \ref{fig:rth}a.}

\textcolor{black}{We find that the {}``steering'' of oscillator
$m2$ by measurements on oscillator $m1$ is very sensitive to the
noise $n_{m2}$ on $m2$, and depends logarithmically on $n_{m2}$
in the limit $n_{m2}\rightarrow\infty$, but is }\textcolor{black}{\emph{insensitive}}\textcolor{black}{{}
to the noise $n_{m1}$ on $m1$. For $n_{m2}=0$, $r_{epr}=0$ and
no thermal barrier exists, which gives a similar behaviour to that
of the hybrid paradox $E_{c|m1}$. }We note however that for $n_{m1}=0$,
a thermal barrier does exist and the threshold is given by $r_{epr}\rightarrow\frac{1}{2}\ln{2n_{m2}}$
as $n_{m2}\rightarrow\infty$. This gives a different sort of behaviour
to that of the hybrid paradox $E_{m1|c}$. In that case, the threshold
for the steering of the oscillator (by a noiseless pulse) was fixed
at $r_{epr}=\frac{1}{2}\ln2$ as the thermal noise of the oscillator
increased. \textcolor{black}{In this way, we learn that whether a
thermally insensitive threshold exists for the steering of a mechanical
oscillator will depend on the nature of entanglement preparation.
 }

\textcolor{black}{We can also consider the steering of the oscillator
$m1$ by $m2$, which is the EPR paradox obtained when $E_{m1|m2}<1$.
We find}\textcolor{blue}{{} }
\begin{eqnarray}
\{\Delta\left(X_{m1}^{out}+gX_{m2}^{out}\right)\}^{2} & = & \left[{\color{red}{\color{black}e^{r}-g\left(e^{r}-e^{-r}\right)}}\right]^{2}\{\Delta X_{m1}^{in}\}^{2}\nonumber \\
 &  & +(1-g)^{2}\left(e^{2r}-1\right)\{\Delta P_{c}^{in}\}^{2}\nonumber \\
 &  & +g^{2}e^{-2r}\{\Delta X_{m2}^{in}\}^{2},\label{eq:sol2}
\end{eqnarray}
and $\Delta_{inf}P_{m1}^{out}=\Delta\left(P_{m1}^{out}-gP_{m2}^{out}\right)=\Delta\left(X_{m1}^{out}+gX_{m2}^{out}\right)=\Delta_{inf}X_{m1}^{out}$.
The optimal $g$ for the detection of $E_{m1|m2}<1$ is given by $\frac{\partial\Delta_{inf}X_{m1}^{out}}{\partial g}=0$.
Solving gives
\begin{eqnarray}
g & = & \left(e^{2r}-1\right)\left(\{\Delta X_{m1}^{in}\}^{2}+\{\Delta P_{c}^{in}\}^{2}\right)\Bigl/\Bigl\{(e^{2r}+e^{-2r}-2)\nonumber \\
 &  & \times\{\Delta X_{m1}^{in}\}^{2}+\left(e^{2r}-1\right)\{\Delta P_{c}^{in}\}^{2}+e^{-2r}\{\Delta X_{m2}^{in}\}^{2}\Bigl\}\nonumber \\
\label{eq:gepr}
\end{eqnarray}
In this case, a thermally insensitive barrier to steering does exist
i.e. the threshold for the steering of oscillator $m1$ becomes insensitive
to $n_{m1}$, as $n_{m1}\rightarrow\infty$. In fact, as $n_{m1}\rightarrow\infty$,
\begin{align}
r_{epr} & =\frac{1}{2}ln(n_{m2}+2-\frac{1}{2(1+n_{m1})}\nonumber \\
 & +\sqrt{\left[n_{m2}+2-\frac{1}{2(1+n_{m1})}\right]^{2}-\frac{2n_{m2}}{1+n_{m1}}-2})\nonumber \\
 & \rightarrow\frac{1}{2}ln2n_{m2}
\end{align}
as evident in the plots of Fig. \ref{fig:rth}b and c. \textcolor{red}{}For
$n_{m2}=0$,
\begin{align}
r_{epr} & =\frac{1}{2}ln[2-\frac{1}{2(1+n_{m1})}+\sqrt{\left[2-\frac{1}{2(1+n_{m1})}\right]^{2}-2}],\nonumber \\
\label{eq:repr4}
\end{align}
which approaches a fixed value as $n_{m1}\rightarrow\infty$, consistent
with the result for the threshold $r_{epr}$ for the steering of the
oscillator $m1$ by the pulse. However, we note now (different to
the result for $E_{m2|m1}$) that there is a sensitivity to the noise
$n_{m2}$ of the {}``steering'' system. The limiting value of $r_{epr}$
increases with $n_{m2}$. For $n_{m1}=0$, the optimal $g$ given
by Eq. (\ref{eq:gepr}) corresponds to a threshold squeezing parameter
of $r_{epr}=\frac{1}{2}\ln2n_{m2}$ in the limit of large $n_{m2}$.\textcolor{red}{{} }

In short, the steering between the two oscillators will become limited
by the thermal noise on them. With a certain method of entanglement
preparation, the steering threshold depends logarithmically on the
thermal noise of the system being steered. In this case, there is
very little dependence on the thermal noise of the steering system.
If the entanglement is prepared another way, an oscillator can be
steered (in the large thermal limit) independently of its own thermal
noise, but then the steering threshold becomes sensitive logarithmically
to large levels of thermal noise on the steering system.\textcolor{red}{}

If we consider equal thermal noise levels $n_{0}$ for the two oscillators,
the thermal barrier for the quantum steering (of either oscillator)
remains sensitive to the thermal noise $n_{0}$ in the limit of large
$r$: the threshold becomes $r>\frac{1}{2}\ln4n_{0}$. This tells
us that enough thermal noise will destroy the possibility of an EPR
paradox, for any given squeeze parameter $r$ that creates the entanglement. 

The interesting feature noticed for this method of entanglement generation
is that both the steering of oscillator $m2$ by $m1$, and the steering
of $m1$ by $m2$, show an insensitivity to the thermal excitation
level $n_{1}$ of oscillator $m1$. Thus, the {}``steering'' of
an oscillator $M$ by another $S$ can be largely insensitive to the
thermal excitation of $M$, or largely insensitive to the excitation
of $S$, depending on the method of entanglement. \textcolor{red}{}\textcolor{red}{}

\section{Conclusion}

In summary, we have examined the effect of an initial thermal excitation
of an oscillator on observing an EPR paradox between a mesoscopic
mechanical oscillator and a pulse. A thermal barrier exists for an
EPR paradox that can demonstrate a quantum {}``steering'' of the
mechanical oscillator. This is the most interesting paradox, since
it tests local reality for a massive, mesoscopic system.

No equivalent thermal barrier exists for the EPR paradox that can
demonstrate a {}``steering'' of the optical pulse. Similarly, as
must be the case given that all types of steering require entanglement,
no barrier exists for the entanglement between the oscillator and
the pulse. This robust pulse-oscillator entanglement is only detectable
using fully sensitive entanglement criteria, such as obtained by the
positive partial transpose PPT method.

Importantly, the thermal barrier to the steering of the oscillator
by the pulse is not insurmountable: it can be overcome for a large
enough squeezing parameter. For temperatures above $100mK$, the condition
is $r>0.4$. This is much more favourable than the conditions $r>2.4$
at $T\sim100mK$, and $r>7$ at $T=293K$, required for observation
of the symmetrically measured entanglement (where $g=1$) \cite{eprentosc}. 

Our results show that the thermal noise of the mechanical object destroys
the manifestation of that paradox, and that the thermal barrier increases
with initial thermal excitation number $n_{0}$ if we consider the
paradox between two symmetric oscillators. This gives an explanation
of the difficulty of observing mesoscopic EPR paradox effects between
massive oscillators at room temperature.

We show that a very big advantage is to be gained if we consider the
EPR paradox of an oscillator with an optical field, which is not thermally
excited, since then (for this simple model) the threshold interaction
to show quantum steering of the oscillator becomes fixed for $n_{0}\gg1$.
By analysing an entanglement swapping scenario that leads to an entanglement
of two thermally excited oscillators, we show that this advantage
is lost for the EPR paradox between two symmetric oscillators.

We conclude by commenting that a practical prediction for the EPR
paradox must fully incorporate the main sources of decoherence. The
practical limitation is that larger pulse-cavity interaction times
lead to increased mechanical decoherence, due to the coupling to the
environmental heatbath at temperature $T_{bath}$. The results presented
here are based on an idealised model which ignores the coupling to
the heat bath of the environment. More complete models have been given
in Ref. \cite{eprentosc}. Those authors did not however analyse the
effects of an environmental heat bath on the EPR paradox, but restricted
to a study of the symmetric DGCZ entanglement. Nonetheless, they estimated
that the symmetric entanglement is feasible, provided $Qf\gg k_{B}T_{bath}/h$,
where $T_{bath}$ is the temperature of the environment, $f$ is the
frequency of mechanical vibration and $Q$ is the cavity quality.
Based on the results of this paper, more optimistic predictions would
be expected, both for entanglement as detected by the PPT condition,
and also for an EPR paradox. \textbf{\emph{}}
\begin{acknowledgments}
We thank P. Drummond, W. Bowen and S. Hofer for useful discussions
and information. We acknowledge support from the Australian Research
Council via Discovery and DECRA grants. Q. Y. H. thanks the National
Natural Science Foundation of China under Grant No. 11121091 and 11274025.
\end{acknowledgments}

\end{document}